\documentclass[published, twocolumn, 12pt]{aastex63}

\usepackage{xspace}

\def\lsim{~\rlap{$<$}{\lower 1.0ex\hbox{$\sim$}}}
\def\gsim{~\rlap{$>$}{\lower 1.0ex\hbox{$\sim$}}}

\def\revc{Rev\,C\xspace}
\def\revd{Rev\,D\xspace}

\def\smili{\texttt{SMILI}\xspace}
\def\casa{\texttt{CASA}\xspace}

\def\clean{\texttt{CLEAN}\xspace}
\def\msclean{\texttt{MS-CLEAN}\xspace}

\begin{document}

\title{\large Next Generation Very Large Array Memo No. 95:\\ 
Evaluation of the Revision D array configuration for stellar imaging}

\author[0000-0001-9401-3117]{Catherine Petretti}
\affiliation{Department of Astrophysics \& Planetary Science, Villanova University, 800 Lancaster Ave, Villanova, PA, 19085, USA}
\affil{Massachusetts Institute of Technology Haystack Observatory, 99 Millstone Road, Westford, MA 01886, USA}

\author[0000-0002-9475-4254]{Kazunori Akiyama}
\affiliation{Massachusetts Institute of Technology Haystack Observatory, 99 Millstone Road, Westford, MA 01886, USA}
\affil{National Astronomical Observatory of Japan, 2-21-1 Osawa, Mitaka, Tokyo 181-8588, Japan}
\affil{Black Hole Initiative, Harvard University, 20 Garden Street, Cambridge, MA 02138, USA}

\author[0000-0002-3728-8082]{Lynn D. Matthews}
\affiliation{Massachusetts Institute of Technology Haystack Observatory, 99 Millstone Road, Westford, MA 01886, USA}

\begin{abstract}
A transformative science case for the proposed next-generation Very Large Array (ngVLA) is resolving the surfaces of nearby stars, both spatially and temporally, enabled by the combination of milliarcsecond-scale resolution and unprecedented sensitivity to thermal radio emission. In a previous study, we demonstrated the feasibility of stellar imaging with simulated observations of nearby stars, using both traditional \clean techniques and newly developed regularized maximum likelihood (RML) imaging methods for image reconstruction.
In this memo, we present a continued study of stellar imaging with the ngVLA, evaluating the imaging capability of the Revision D (henceforth \revd) Main Array configuration compared to the previous Revision C (henceforth \revc) configuration.
We find that the \revd configuration, with more uniform coverage and better circular symmetry, improves the synthesized beam, resulting in better \clean reconstructions of simulated images of evolved stars with complex morphology, especially with robust weighting.
However, the highly non-Gaussian nature of the synthesized beam still persists with both robust and natural weightings in the \revd configuration and continues to limit the image fidelity of image reconstructions with non-uniform weighting.
The RML methods show stable performance that is resilient to different array configurations with image quality comparable to or better than \clean methods in the presented simulation, consistent with our previous work.
Our simulation results suggest that the \revd configuration will provide a better deconvolution beam compared with the \revc configuration, which would enhance the imaging capability for non-uniform weighting, and they continue to demonstrate that RML methods are an attractive choice, even for the improved array configuration.
\vspace{4eM} 
\end{abstract}

\section{Introduction\protect\label{intro}}
The next-generation Very Large Array (ngVLA) is currently planned to be a heterogeneous array of 244 antennas of 18~m diameter and 19 dishes with 6~m diameter \citep{Selina2018}, with the array design based on designated ``key science goals'' \citep{Murphy2018}.
To meet diverse scientific needs, the planned ngVLA array is designed to have three subarrays: a ``Short Baseline Array'' with baselines of 11--56~m, a ``Main Array'' with 214 of the 18~m antennas on baselines ranging from tens of meters to $\sim$1000~km, and a ``Long Baseline Array'' with 30 of the 18~m antennas spread across the North American continent for Very Long Baseline Interferometry (VLBI).

The ngVLA Main Array will be able to detect thermal emission at milliarcsecond-scale resolution. Until now, such angular scales have only been accessible with VLBI for compact, non-thermal objects with a high brightness temperature.
A groundbreaking scientific application of the Main Array will be its ability to obtain resolved images of the surfaces of nearby stars spanning a range of spectral types and evolutionary phases from dwarfs to supergiants \citep{Carilli+2018,MattClaussen2018, Harper2018, Akiyama2019}.

The current ngVLA design has a Main Array that is ``tri-scaled'' \citep[e.g.][]{Carilli2017,Carilli2018}, comprising: (1) a densely sampled, 1~km-diameter core of 94 antennas; (2) a VLA-scale array of 74 antennas with baselines up to  $\sim$30~km; and (3) extended baselines (46 stations) out to $\sim$1000~km. 
Although the Main Array is suited to meeting the ngVLA's combined requirements for angular resolution, point source sensitivity,
and surface brightness sensitivity, its antenna distribution results in a highly non-Gaussian synthesized beam.
The beam shape comprises a narrow core and a two-tiered ``skirt,'' or long tail of side lobes \citep[][]{Carilli2017, Carilli2018}.
This poses a challenge for imaging ngVLA data with traditional \clean deconvolution methods, in which a model of the ideal ``\clean beam'' is determined by fitting a Gaussian to the dirty beam point spread function \citep[e.g.,][]{Hogbom74}.
A consequence is that it is difficult to achieve maximum angular resolution in an ngVLA \clean image without sacrificing sensitivity \citep{Carilli2017, Carilli2018, Rosero2019}. 
This issue cannot be overcome through the use of robust weighting \citep{Briggs99} during the deconvolution \citep{Carilli2017}, and it therefore currently presents a potential inherent limitation to the array performance. 

\begin{figure*}[t]
    \centering
    \gridline{
        \fig{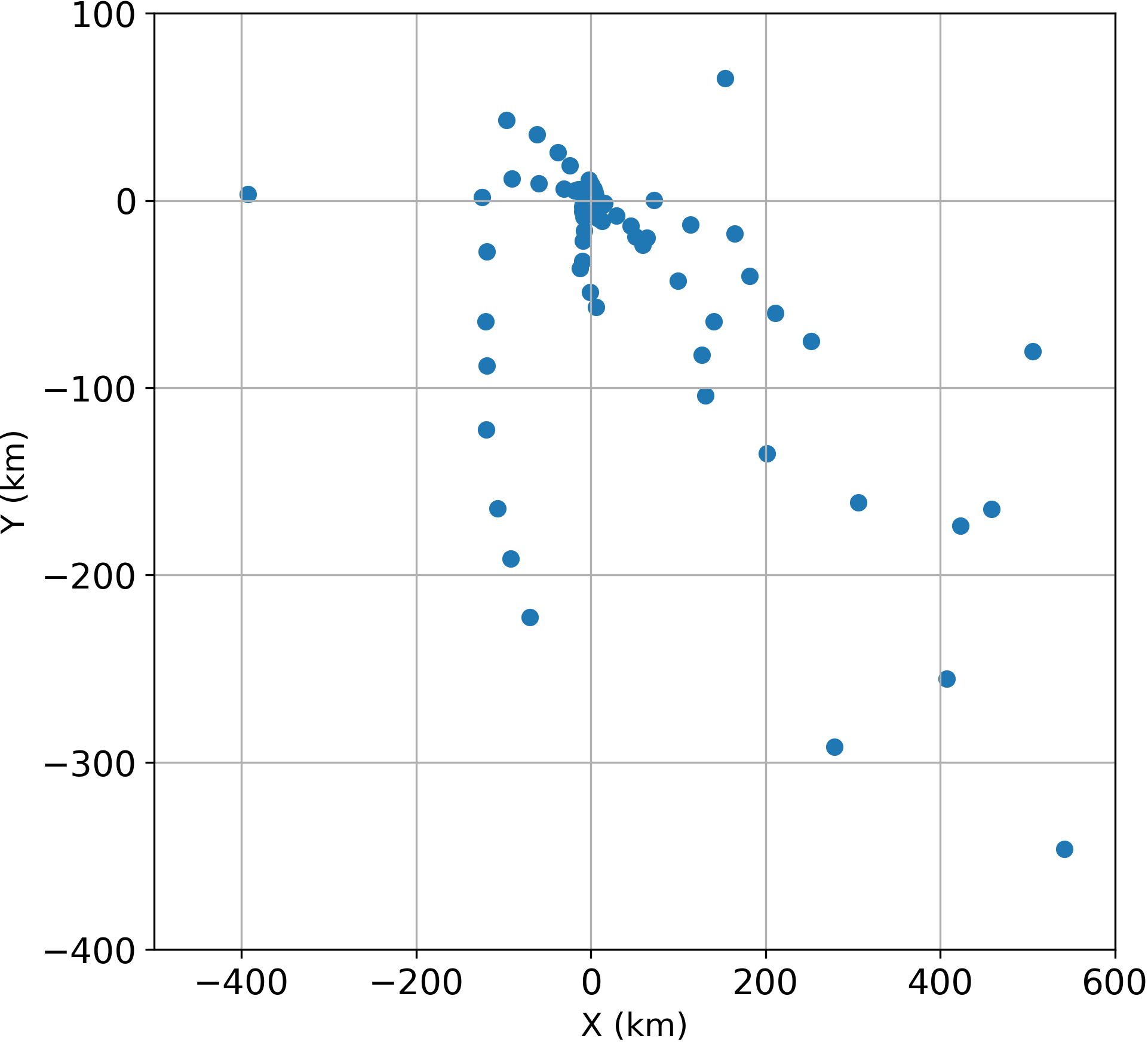}{0.4\textwidth}{(a) \revc Configuration}
        \fig{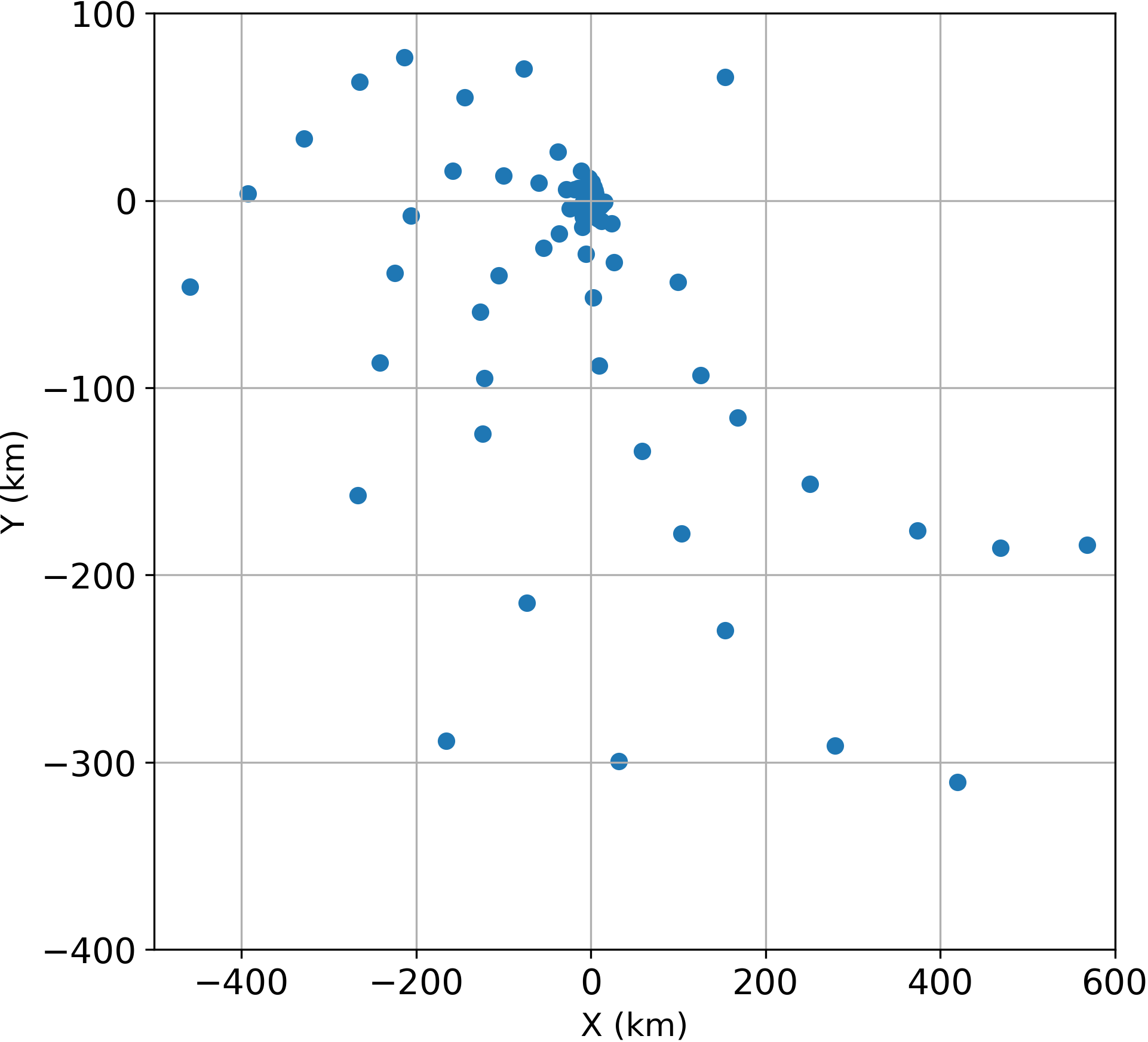}{0.4\textwidth}{(b) \revd Configuration}
    }
    \caption{Baseline configurations of (a) \revc and (b) \revd Main Arrays for the ngVLA.}
    \label{fig:configs}
\end{figure*}

\subsection{Our previous study}
In  a previous ngVLA study \citep[ngVLA  memo  No.~66;][]{Akiyama2019},  we  investigated the application of regularized maximum likelihood (RML) methods \citep[see][for an overview]{EHTC2019d}, a new class of imaging techniques developed for the Event Horizon Telescope, to simulate ngVLA Main Array observations of stellar radio photospheres as a test case. 
RML methods take a forward-modeling approach, directly solving for the images without using either the dirty beam or the dirty map. 
Consequently, this method has the potential to improve the fidelity and effective angular resolution of ngVLA images in three ways: (1) allowing high-fidelity reconstructions, even at modest super resolution 2-3 times finer than that of traditional \clean \citep[e.g.][]{Honma2014, Chael2016, Akiyama2017a, Akiyama2017b, Kuramochi2018}; (2) the capability to reconstruct images directly from closure quantities, free from antenna-based calibration errors \citep[e.g.][]{Chael2016, Chael2018,Akiyama2017b}; and (3) the ability to handle intrinsically multi-dimensional emission, such as time-variable emission structures \citep[e.g.][]{Johnson2017}. The aforementioned advantages of RML methods over \clean have now been demonstrated not only for VLBI but also for VLA and ALMA continuum observations \citep{Matthews2018, Chael2018, Yamaguchi2020}.

We found that both Multi-scale \clean \citep[henceforth \msclean;][]{Cornwell2008} and RML methods can provide high-fidelity images recovering most of the representative structures for different types of stellar photosphere models imaged with the ngVLA. 
However, RML methods show better performance than \msclean for various stellar models in terms of goodness-of-fit to the data, residual errors of the images, and recovering representative features in the groundtruth images.
Our simulations support the feasibility of transformative stellar imaging science with the ngVLA and simultaneously demonstrate that RML methods are an attractive choice for ngVLA imaging.

\subsection{Revision D array configuration and scope of this memo\protect\label{baselines}}
An updated planned configuration of the ngVLA Main Array, Revision D (hereafter \revd), was recently developed by the ngVLA Imaging and Calibration Working Group as part of Conceptual Design Review (CoDR) preparation \citep{Carilli+2021}. 
The \revd configuration has many updates from the previous configuration, revision C.01\footnote{available in https://ngvla.nrao.edu/page/tools} (henceforth \revc) adopted in our previous memo \citep{Akiyama2019}. 
The major changes in the Main Array relative to the \revc configuration include: (1) the 46 antennas for extended baselines from $\sim$30\,km to $\sim$1000\,km are more distributed; and (2) the central core array with 94 antennas has been expanded substantially from $\sim$1\,km to $\sim$4\,km.

The updated \revd design exhibits substantial changes in the antenna positions from the previous revision. In Figure \ref{fig:configs}, we show the antenna positions of both the \revc and \revd configurations. The \revd design has five ``spiral arms" of antennas, providing extended baseline lengths ranging from $\sim$30\,km to $\sim$1000\,km, whose shapes are much clearer than the \revc design. The antenna positions are more evenly distributed both along each arm and between arms, which may provide substantial improvements in $uv$-coverage (see Section~\ref{sims}).

In this memo we continue the work of \citet{Akiyama2019} by evaluating the imaging capability of the \revd configuration for stellar imaging. We performed imaging simulations of several stellar photosphere models using \msclean and RML methods with both \revc and \revd configurations to assess the improvement with the updated array configuration. In \clean imaging, we also explore the performance of different $uv$-weightings, whereas our previous work only explored uniform weighting.

\section{Models and Simulated Observations\protect\label{modelsandsims}}

\subsection{Models\protect\label{models}}
We evaluated the imaging performance for both the \revc and \revd configurations using a series of four different simulated data sets, adopted in our previous study \citep{Akiyama2019}. Here, we briefly describe each model. For more details, please see \cite{Akiyama2019}.

The first series of models comprises simple geometric disk models based on a uniform disk brightness distribution with surface features superposed. This class of models is motivated by the observed brightness distribution of the radio photospheres of nearby asymptotic giant branch (AGB) and red supergiant (RSG) stars, whose radio emission are well represented by a uniform disk (either circular or elliptical) at the angular resolutions of the current VLA and ALMA ($\sim$20--40\,mas), which is sufficient only to marginally resolve the radio photospheres of the nearest AGB and RSG stars \citep[e.g.,][]{Lim1998, RM97, RM07, Menten2012, Matthews2015, Matt+2018, Vlemmings2017}. We adopted two models created in our previous memo, \texttt{UniDisk222pc} and \texttt{UniDisk1kpc}, with uniform circular brightness distributions and three ``spots" of different sizes superposed (one brighter than the underlying photosphere and two that are cooler). \texttt{UniDisk222pc} has a uniform (circular) disk diameter of 80~mas and a flux density at 46.1~GHz of 28.0~mJy, similar to those of the RSG star Betelgeuse as measured with the VLA at 7~mm \citep{Lim1998}.
\texttt{UniDisk1kpc} is an additional version appropriately scaled to a distance of 1~kpc. Both sources are located at the J2000 sky coordinates of RA=02$^{\rm h}$ 00$^{\rm m}$, DEC=$-02^{\circ}$ 00$'$, which were intentionally chosen to result in a slightly elliptical dirty beam.

The second and third series of models we explored are based on more physically-motivated images created by 3D hydrodynamic simulations of AGB and RSG star atmospheres from \cite{Freytag2017} and \cite{Chiavassa2009}. Our adaptions of these models are referred to as the \texttt{Freytag} and \texttt{Chiavassa} models, respectively, in our previous memo \citep{Akiyama2019}. 
These models are based on near-infrared images from 3D hydrodynamic simulations since no detailed, high-resolution models are currently available that predict the appearances of AGB or RSG stars at millimeter wavelengths. 
However, our adoption of the near-infrared models as morphological templates for the radio emission is motivated by the growing evidence based on recent VLA and ALMA imaging that radio photospheres are time-variable and non-uniform in surface brightness \citep[e.g.,][]{OGorman2015, Matthews2015,  Matt+2018, Vlemmings2019}, and the origins of these behaviors may be intricately linked with those that give rise to the complex and time-varying appearance of the star at near-infrared and shorter wavelengths \citep[e.g.,][]{Matt+2018}. 
While the detailed structures of radio photospheres at millimeter wavelengths are poorly constrained, the imaging simulations with these models allow challenging tests for the anticipated stochastic and complex morphology of the brightness distributions.

The \texttt{Freytag} model, simulating the time-variable images of a 1~$M_{\odot}$ AGB star, is based on model st28gm06n25 from \cite{Freytag2017}, which has a bolometric luminosity $L$=6890~$L_{\odot}$, a mean effective temperature $T_{\rm eff}$=2727~K, and a pulsation period $P$=1.388~yr. The location of the star is assumed at a J2000 sky position of RA=02$^{\rm h}$ 19$^{\rm m}$, DEC=$-02^{\circ}$ 58$'$ and a distance of $\sim$150~pc. We scaled the images to subtend a mean angular diameter of $\sim$50~mas and have an integrated flux density of 10~mJy at 46.1\,GHz.

The \texttt{Chiavassa} model, simulating the time-varying appearance of a 12~$M_{\odot}$ RSG star, is based on the $H$-band model st35gm03n07 in \cite{Chiavassa2009} with a bolometric luminosity $L$=93,000~$L_{\odot}$, a mean effective temperature $T_{\rm eff}$=3490~K, and a radius $R$=832~$R_{\odot}$. We adapt this model to represent a radio photosphere whose angular diameter and flux density at 46.1~GHz are $\sim$80~mas and 28~mJy, respectively, comparable to the RSG Betelgeuse, which lies at a distance of $\sim$222~pc \citep{Lim1998}. The star was located at a J2000 sky position also comparable to Betelgeuse (RA=05$^{\rm h}$ 55$^{\rm m}$, DEC=$+07^{\circ}$ 24$'$).

\begin{figure*}[t]
    \centering
    \gridline{
        \fig{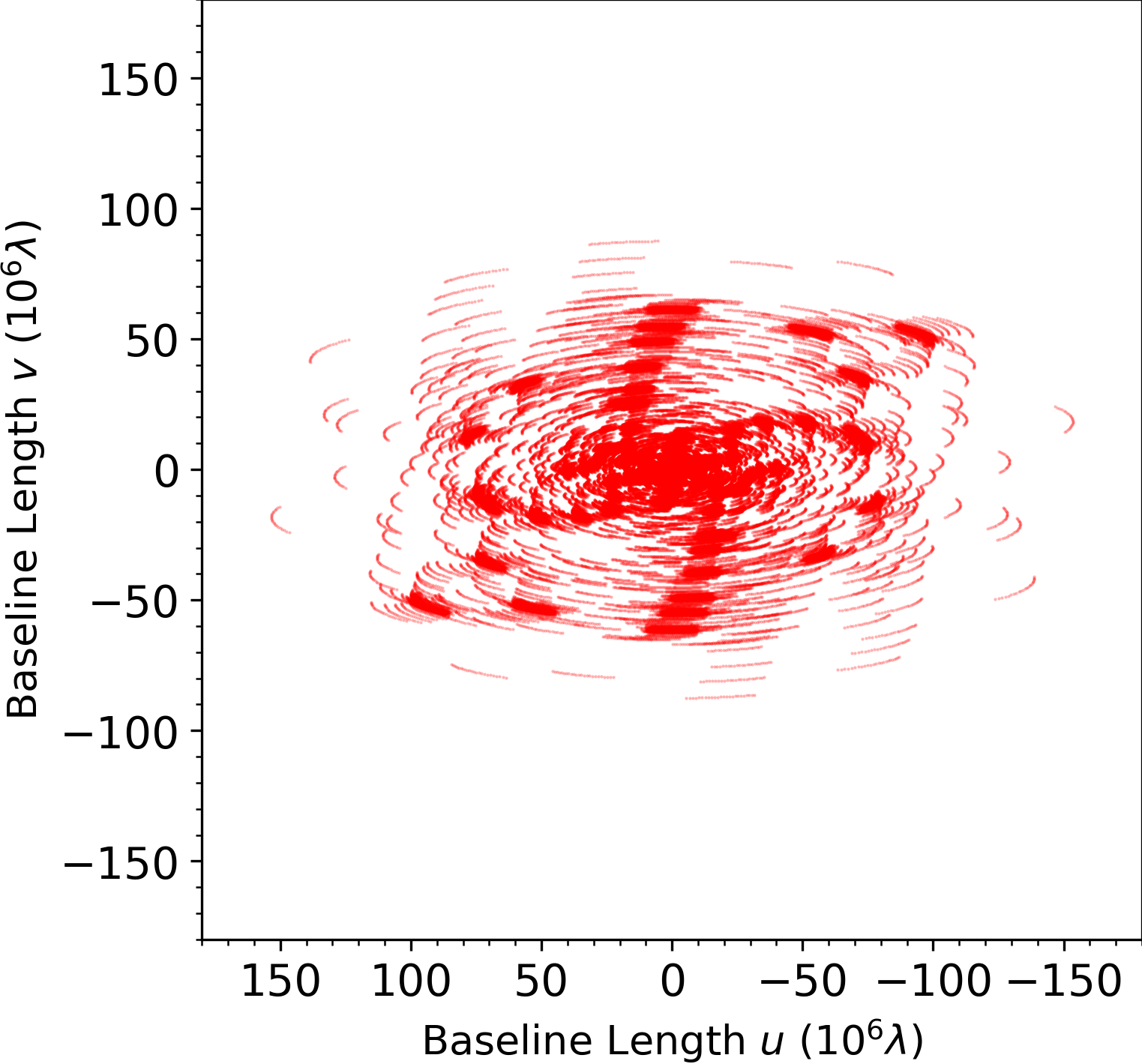}{0.4\textwidth}{(a) \revc Configuration}
        \fig{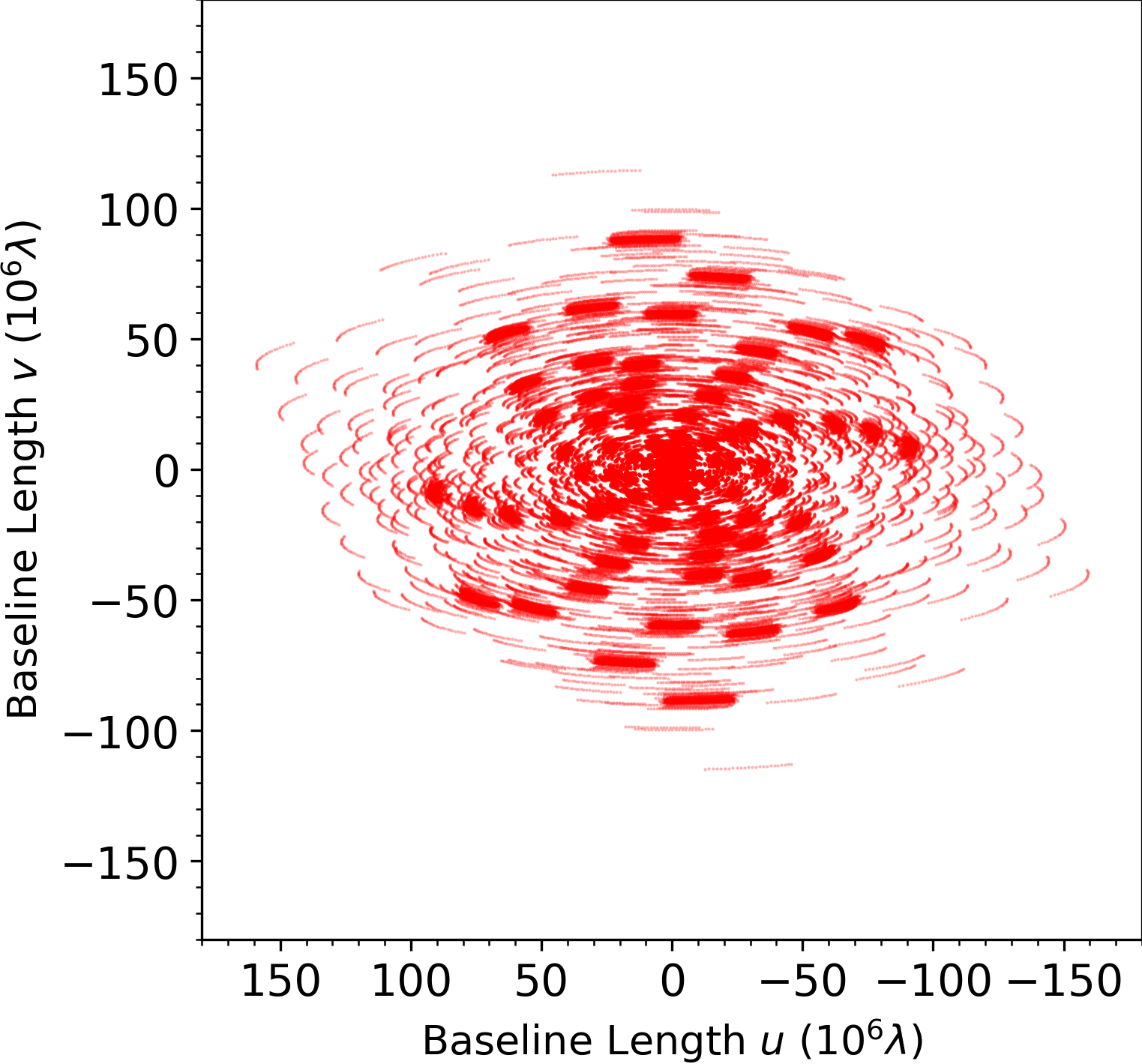}{0.4\textwidth}{(b) \revd Configuration}
    }
    \caption{Example $uv$-coverages of the {\tt Chiavassa} model at 46\,GHz ($\lambda\approx$7~mm). The source is located at a declination of $\sim +7^\circ$ (see Section \ref{modelsandsims} for details). Panels (a) and (b) show $uv$-coverages of the \revc and \revd ngVLA Main Array configurations, respectively.}
    \label{fig:uvcoverage}
\end{figure*}

\subsection{Simulations\protect\label{sims}}
We performed simulated ngVLA observations of each model at 46~GHz using the \texttt{simobserve} task in \texttt{CASA} to produce model visibility data in measurement set (MS) or uvfits format. In all cases, the array configurations were taken to be the ngVLA Main Array (see Section~\ref{intro}) with telescope locations of the \revc and \revd configurations (see Section~\ref{baselines} and Figure~\ref{fig:configs}).

All groundtruth model images have a pixel size of 0.04~mas, a factor of $\sim$25 times smaller than the angular resolution of the ngVLA Main Array at 46~GHz. To avoid edge effects, zero padding was used to create a field-of-view for each groundtruth frame of $\sim$0.33 arcsec per side. 

In this memo, we only consider thermal noise, which was added using the prescription outlined in \cite{Carilli+2017a}. All of our simulations assumed dual polarizations (resulting in Stokes~I images) and a center observing frequency of 46.1~GHz ($\lambda{\approx}$7~mm), allowing direct comparisons with both real and simulated observations from the current VLA. For noise calculation purposes, we assume a total bandwidth of 10~GHz per Stokes \citep[half the nominal value expected for the ngVLA; see][]{Selina2018}. The synthetic observations ranged from 2--4 hours in duration and were assumed be centered on the time of the source transit. 

The resulting $uv$-coverages for the {\tt Chiavassa} model are shown in Figure~\ref{fig:uvcoverage} for both the \revc and \revd configurations. Figure~\ref{fig:uvcoverage} clearly shows that the \revd configuration provides more symmetric $uv$-coverage than \revc. The overall extension of $uv$-coverage is more uniform, which would enhance the circular symmetry of the synthesized beam pattern. Furthermore, the dense clusters of spatial frequencies created by baselines between the central core and outer intermediate/long-baseline antennas are more evenly spread in the \revd configuration. See Section \ref{MSCLEAN} for the resultant synthesized beams at each $uv$-weighting.

\section{Image Reconstructions \protect\label{imaging}}
\begin{figure*}[!t]
    \centering
    \gridline{
        \fig{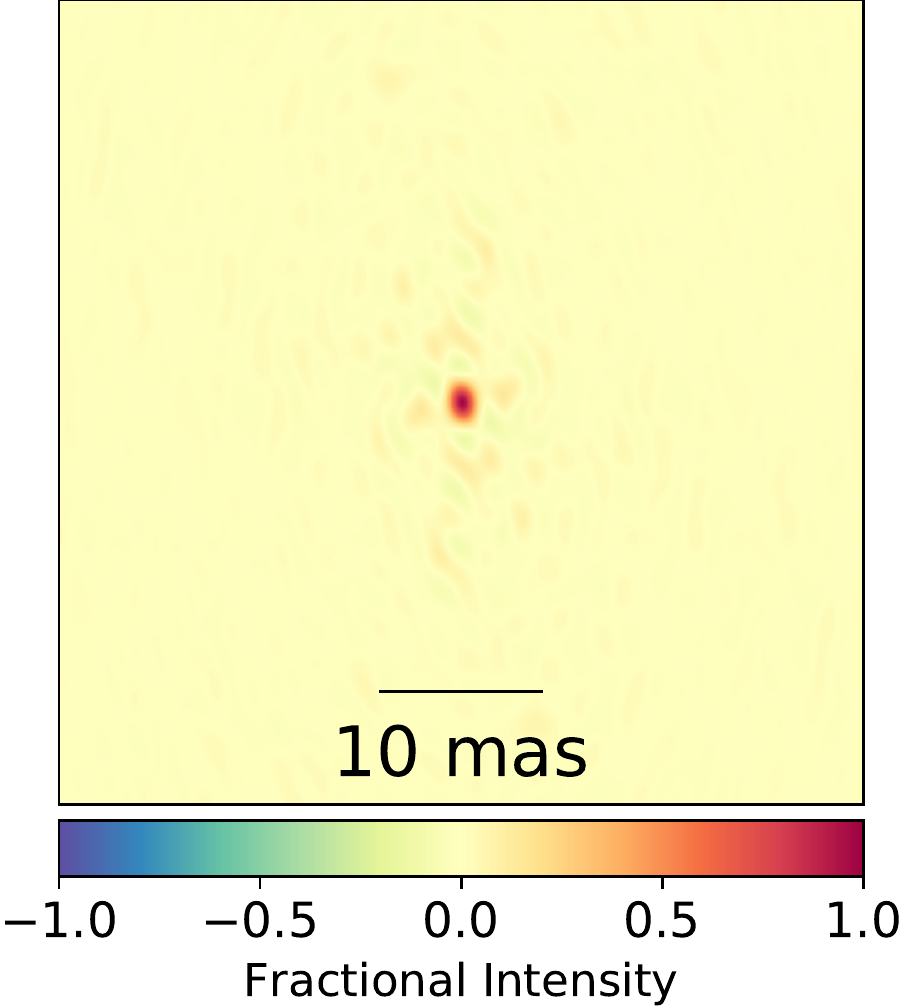}{0.31\textwidth}{(a) \revc Configuration, uniform weighting}
        \fig{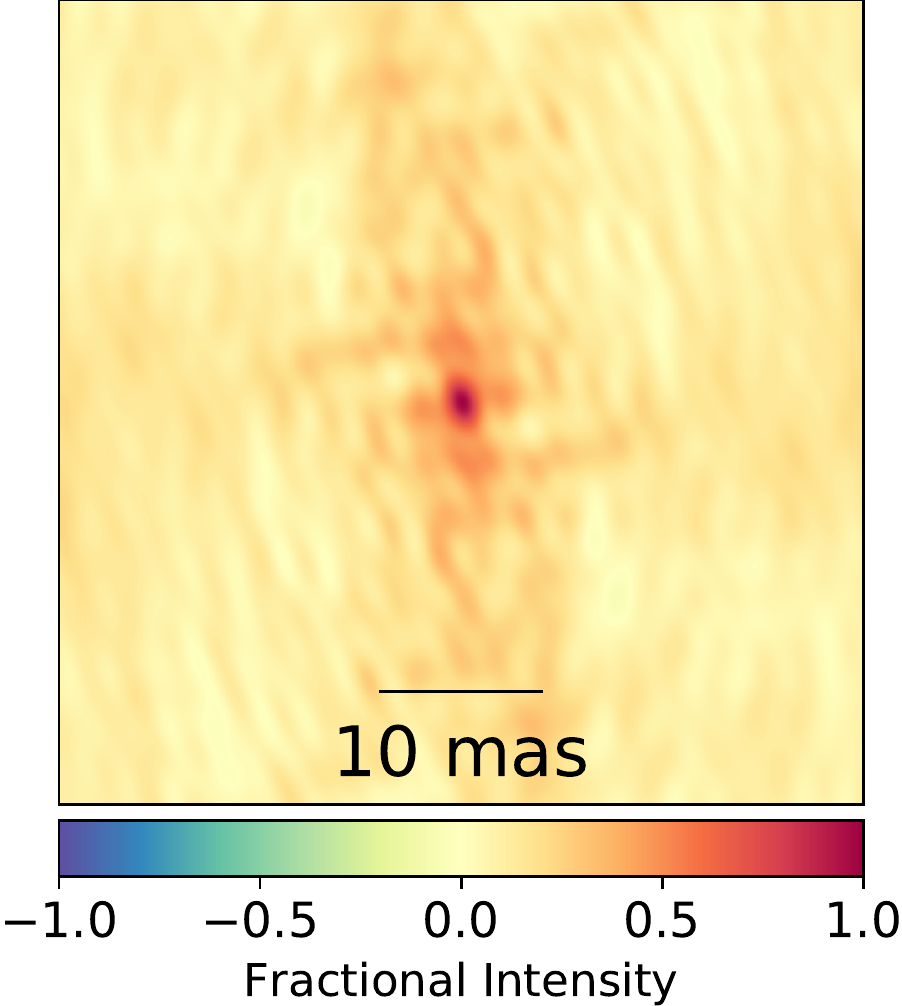}{0.31\textwidth}{(b) \revc Configuration, robust weighting}
        \fig{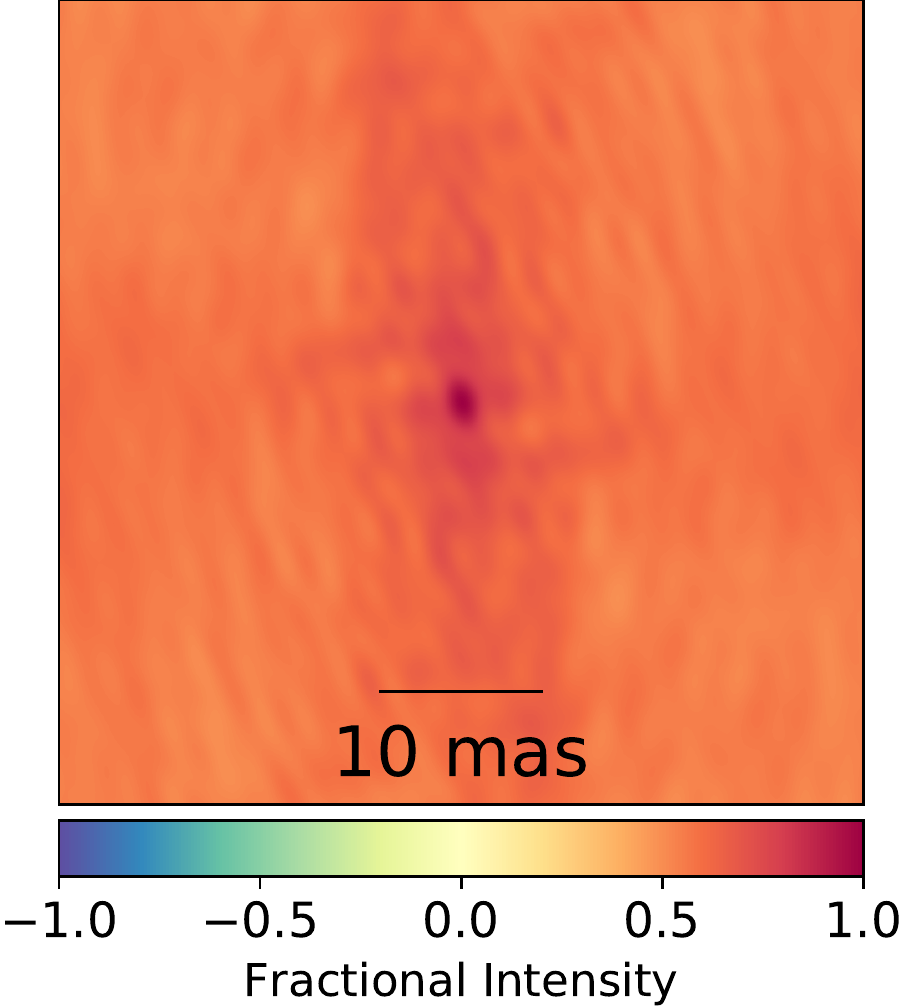}{0.31\textwidth}{(c) \revc Configuration, natural weighting} 
        }
    \gridline{
        \fig{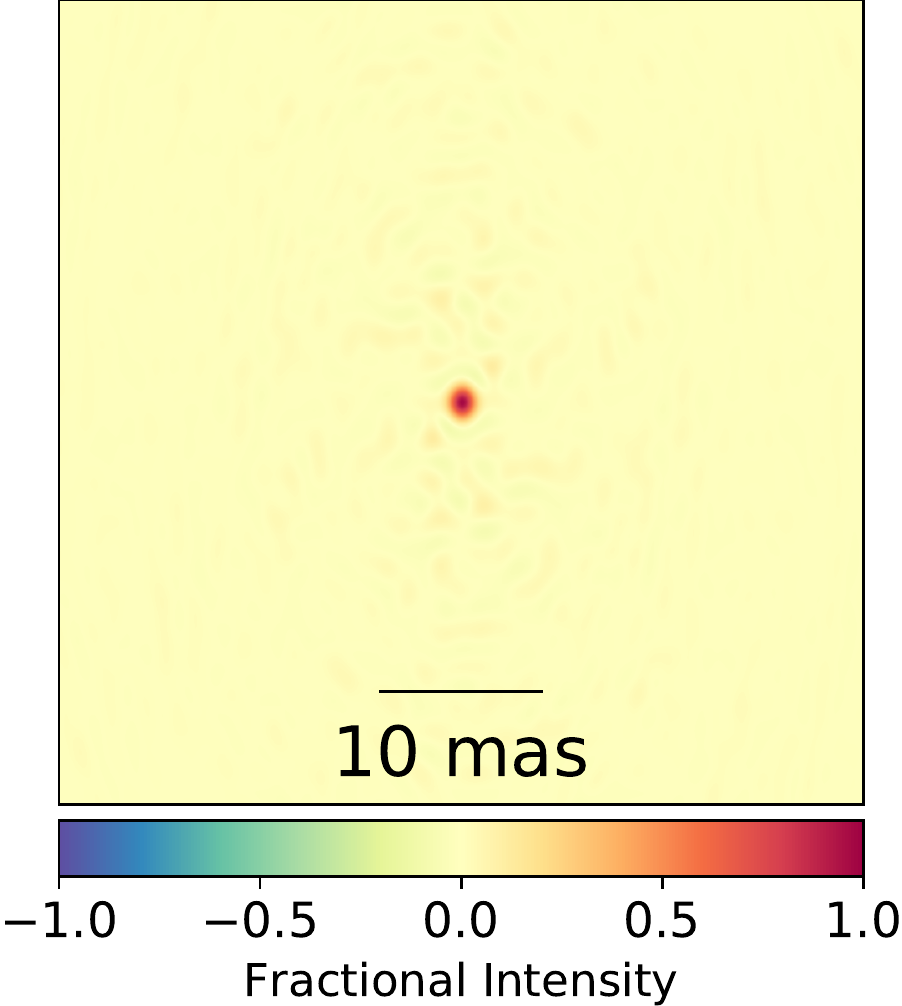}{0.31\textwidth}{(d) \revd Configuration, uniform weighting}
        \fig{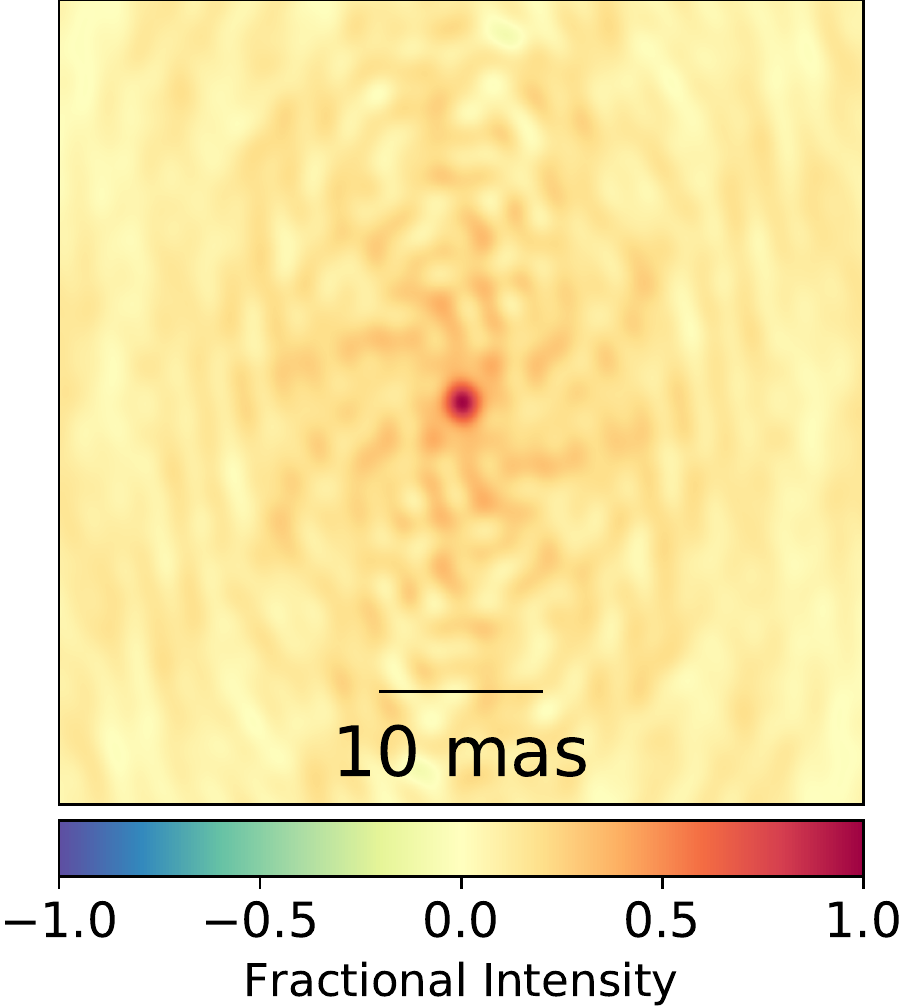}{0.31\textwidth}{(e) \revd Configuration, robust weighting}
        \fig{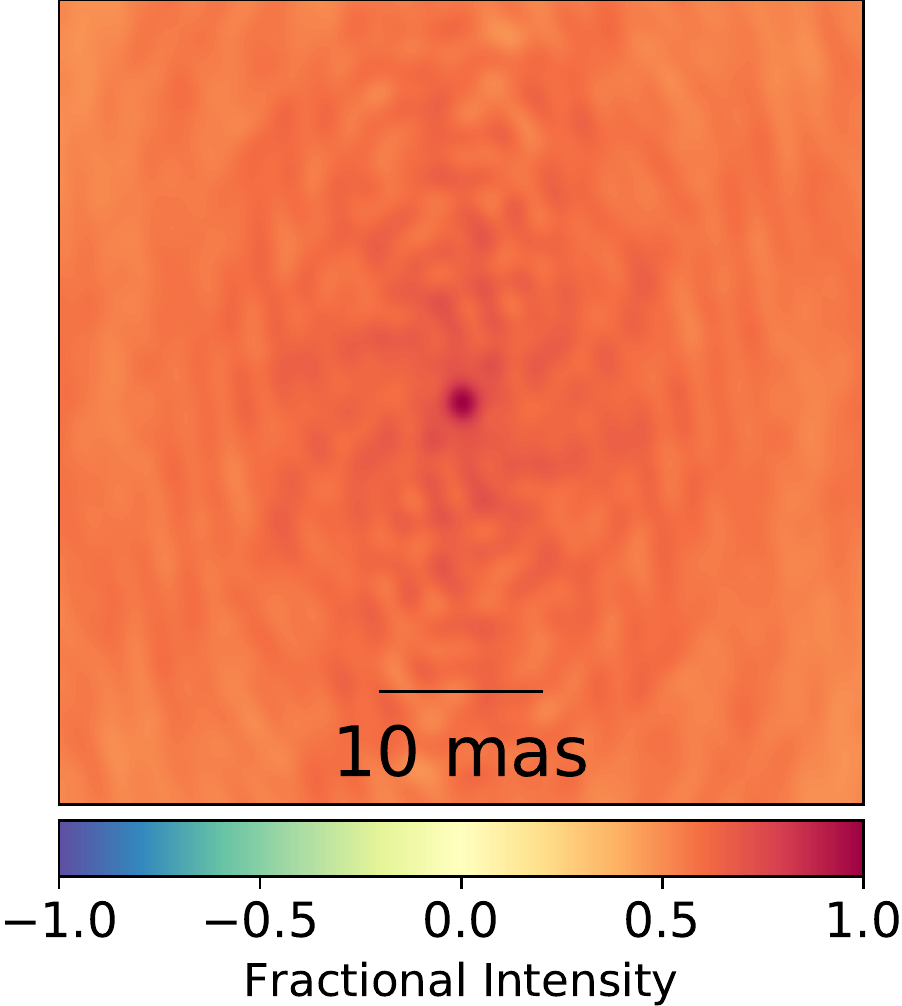}{0.31\textwidth}{(f) \revd Configuration, natural weighting}
    }
    \caption{The synthesized beams at different $uv$-weighting for both array configurations, corresponding to $uv$-coverages shown in Figure~\ref{fig:uvcoverage}. The top panels (a-c) show the synthesized beams for the \revc configuration, while the bottom panels (d-f) show those for the \revd configuration. From the leftmost to rightmost panels, the synthesized beams for uniform, robust and natural weighting, respectively, are shown. The robust parameter for the robust weighting is set to zero to show the intermediate weighting between uniform and natural weightings. See Table~\ref{tab:beam_sizes} for FWHM sizes of each beam and Figure~\ref{fig:beam_slices} for the linear slices of each beam along the RA and Dec axes.}
    \label{fig:beams}
\end{figure*}

\begin{figure*}[!t]
    \centering
    \gridline{
        \fig{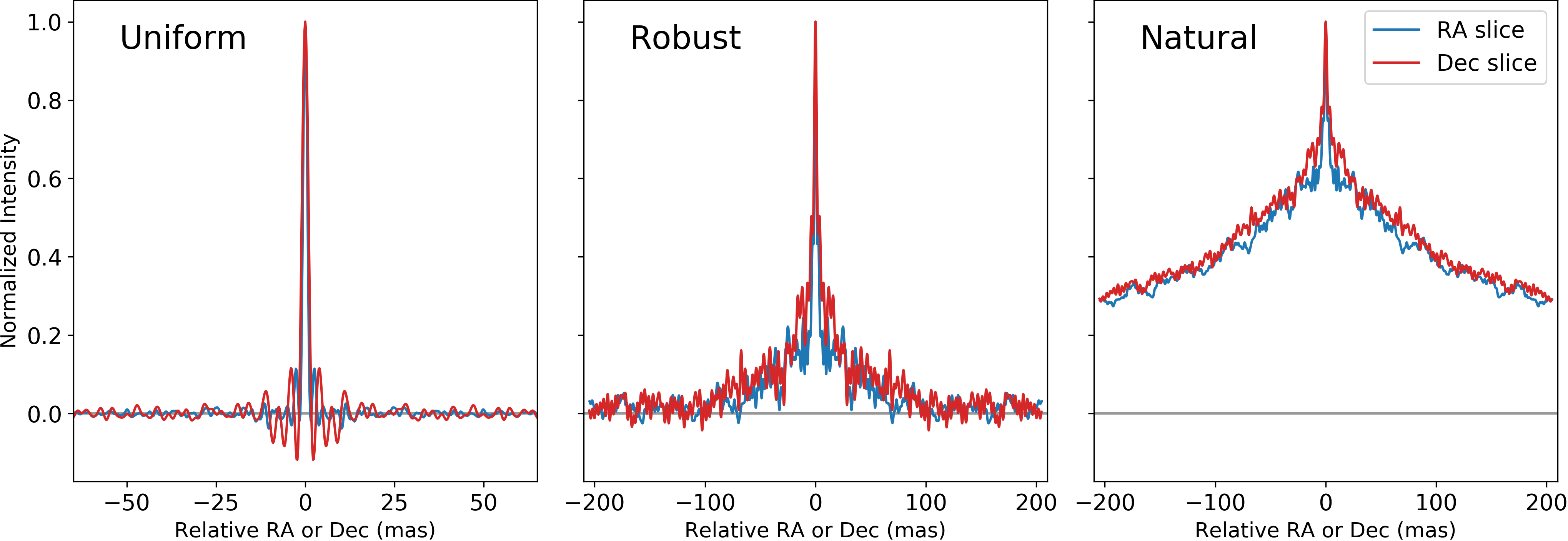}{0.98\textwidth}{(a) \revc Configuration}
        }
    \gridline{
        \fig{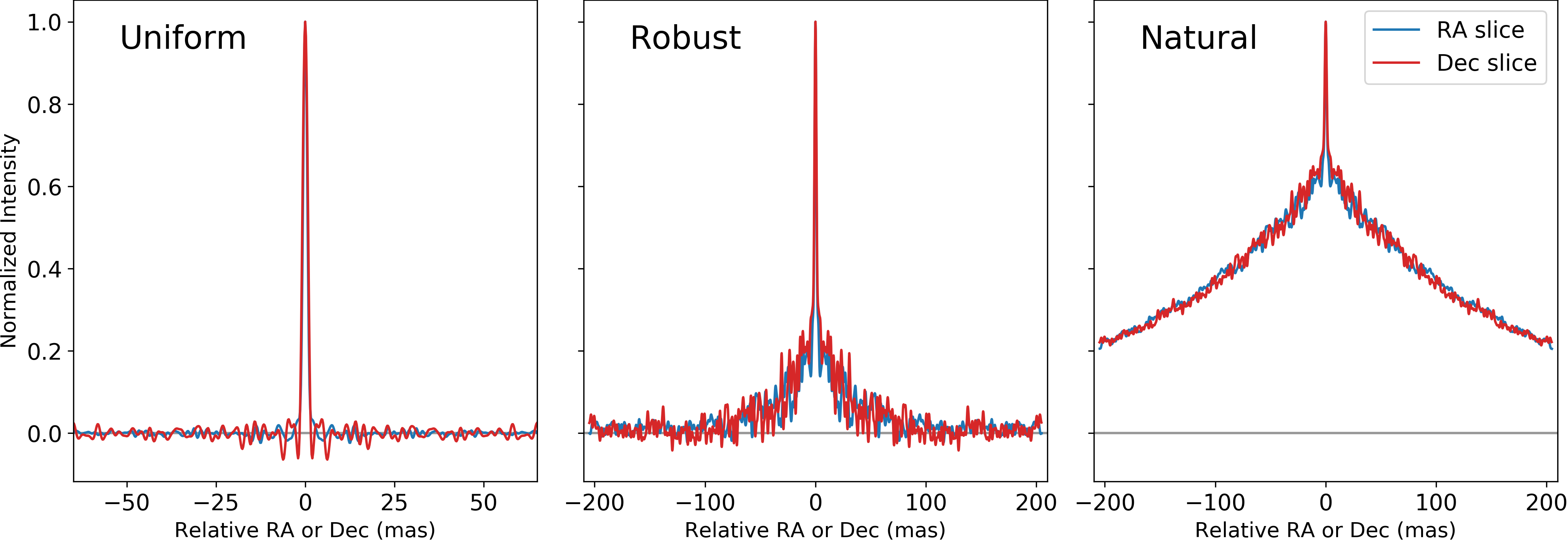}{0.98\textwidth}{(b) \revd Configuration}
        }
    \caption{The linear slices of the synthesized beams shown in Figure~\ref{fig:beams}, along the RA (blue line) and Dec (red line) axes. The top panel (a) shows the slices for the \revc configuration, while the bottom panel (b) shows those for the \revd configuration.}
    \label{fig:beam_slices}
\end{figure*}

\begin{table}[t]
    \centering
    \caption{The parameters of the synthesized beams shown in Figure \ref{fig:beams}.}
    \begin{tabular}{ccccc}
        \hline \hline
        Array Config. & $uv$-weighting & $\theta_{maj}$ & $\theta_{min}$ & $\theta_{PA}$ \\
        & & (mas) & (mas) & ($^\circ$) \\
        \hline
        \revc & Uniform &   2.0 &   1.4 &   7.0 \\
              & Robust  &   7.0 &   3.8 &  10.1 \\
              & Natural &  14.0 &  11.1 &   1.8\\
        \revd & Uniform &   1.7 &   1.4 &  -0.5 \\
              & Robust  &   3.5 &   2.3 & -29.3 \\
              & Natural &  12.0 &  11.2 &   3.4 \\
         \hline
    \end{tabular}
    \label{tab:beam_sizes}
\end{table}

\subsection{RML Imaging}
For our RML imaging investigations, we used \smili\footnote{\url{https://github.com/astrosmili/smili}} \citep[][]{Akiyama2017a, Akiyama2017b}, a Python-interfaced open-source imaging library primarily developed for the Event Horizon Telescope (EHT) that was used in our previous memo \citep{Akiyama2019}. Simulated data (see Section~\ref{modelsandsims}) were exported to uvfits files from \texttt{CASA} and loaded into \texttt{SMILI} for imaging and analysis. Since visibility weights in uvfits files from \texttt{CASA} do not reflect actual thermal noise, they were re-evaluated using the scatter in visibilities within 1~hour blocks using the \texttt{weightcal} method. Images were then reconstructed with full complex visibilities. 

The most relevant parameters for \texttt{SMILI} imaging (or for RML methods more widely) are the pixel size, the field-of-view of the image, and the choice and weights of regularization functions. We adopted a pixel size of 0.2~mas and a field-of-view of 300 pixels for all of the models. For the uniform disk models (\texttt{UniDisk222pc}, \texttt{UniDisk1kpc}) and \texttt{Chiavassa} model, we adopted Total Variation (TV) regularization at the regularization parameter of $10^3$.
TV regularization leads to an edge-preserved smooth image \cite[e.g.][]{Akiyama2017a, Akiyama2017b}, well matched with the anticipated brightness distributions for these models.
For the \texttt{Freytag} model, we adopted a relative entropy term with a flat prior at a regularization parameter of $10^{-4}$. 
This is a classical regularization function used for the Maximum Entropy Method (MEM) known to have a good performance for the edge-smoothed brightness distribution. 
For either of the regularization parameters, a higher value results in stronger regularization on the reconstructed image, leading to a more piece-wise smooth image.

\subsection{Multi-scale CLEAN\protect\label{MSCLEAN}}
For our \clean imaging tests, we used the \texttt{CASA} 5.8.0-109 version of \msclean as implemented via the ``\texttt{tclean}" task. A general overview of \msclean can be found in e.g., \cite{Cornwell2008} \citep[see also][]{Rich2008}. For all \clean images presented, we adopted a loop gain of 0.1, a cell size of 0.2~mas, and 10,000 \clean iterations. No \clean boxes were used. For each model and array configuration, we tested uniform weighting, natural weighting, and robust weighting with a robust parameter of zero to show the intermediate weighting between uniform and natural weightings \citep{Briggs99}.

In Figure~\ref{fig:beams}, we show the synthesized beams for the \texttt{Chiavassa} model for each weighting and array configuration, whose FWHM sizes are summarized in Table~\ref{tab:beam_sizes}. In Figure~\ref{fig:beam_slices}, we show corresponding two-dimensional intensity slices across the beams. 
Regardless of the $uv$-weighting, the \revd configuration provides a more circularly symmetric beam pattern (Figure~\ref{fig:beam_slices}), even for a low declination source, as indicated by $uv$-coverage (Figure~\ref{fig:uvcoverage}).
Furthermore, there are modest improvements in the \revd configuration on the level of the envelope function for side lobes in all $uv$-weightings (Figure~\ref{fig:beam_slices}), brought by the more uniform distributions of the dense clusters of baselines in $uv$-space between the central core and outer antennas. 

Despite the moderate improvement in \revd's synthesized beam, the non-Gaussianity of the beam is persistent in both robust and natural weighting (Figure \ref{fig:beams} and \ref{fig:beam_slices}). The synthesized beams for both of these choices of $uv$-weighting comprise a sharp milliarcsecond-scale primary component superposed on a plateau with long tails extending over scales larger than the primary component, which are consistent with those from earlier Main Array configurations \citep{Carilli2017}.

\section{Results\protect\label{results}}

\begin{figure*}[ht]
    \centering
    \includegraphics[width=1.0\linewidth]{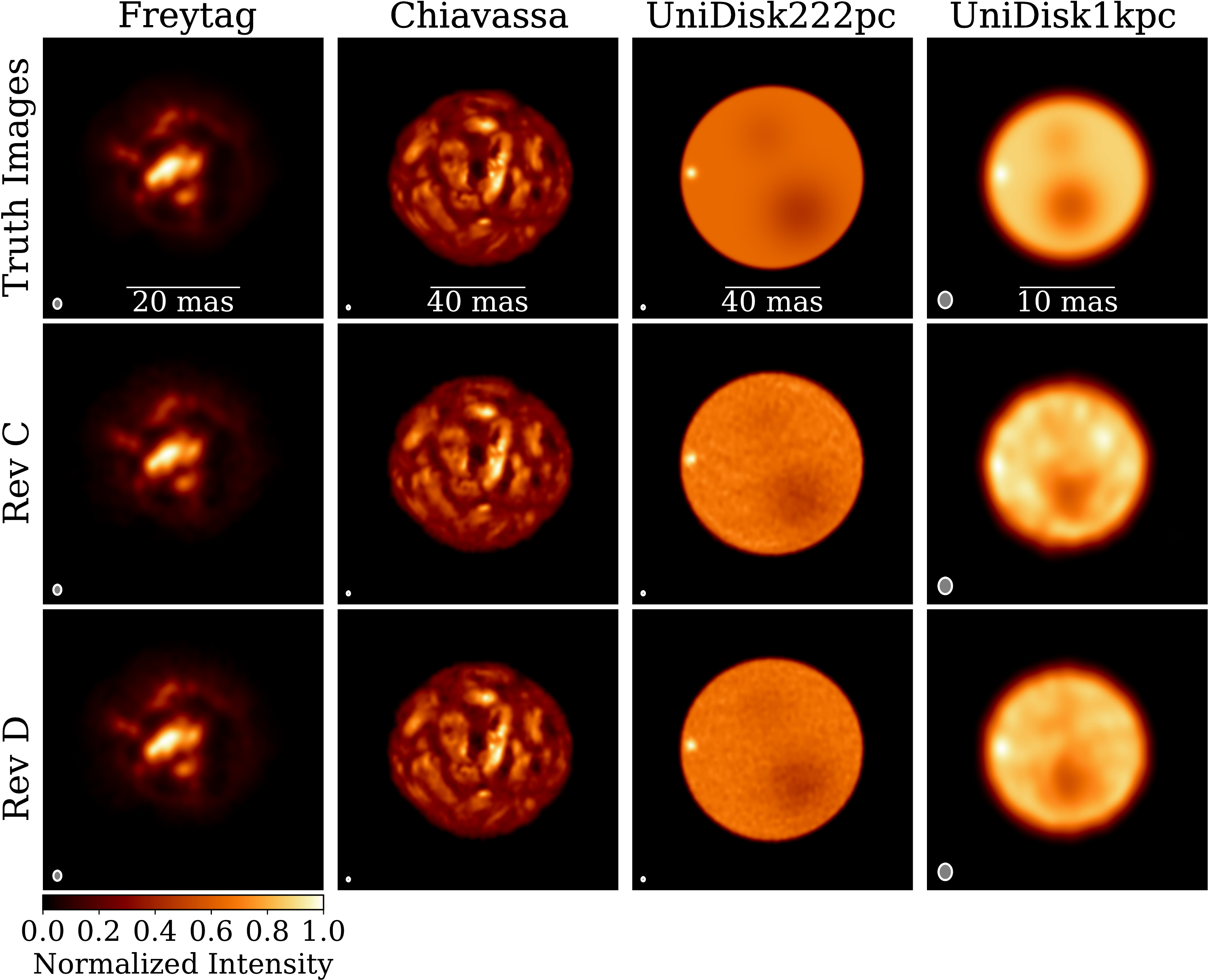}
    \caption{The groundtruth images and corresponding \texttt{SMILI} reconstructions for \revc and \revd configurations for four stellar models. Here, images of each model are blurred at the Gaussian beam for the \revd configuration in uniform weighting used in \msclean reconstructions. }
    \label{fig:rml_imaging}
\end{figure*}

\begin{figure*}[ht]
    \centering
    \gridline{
        \fig{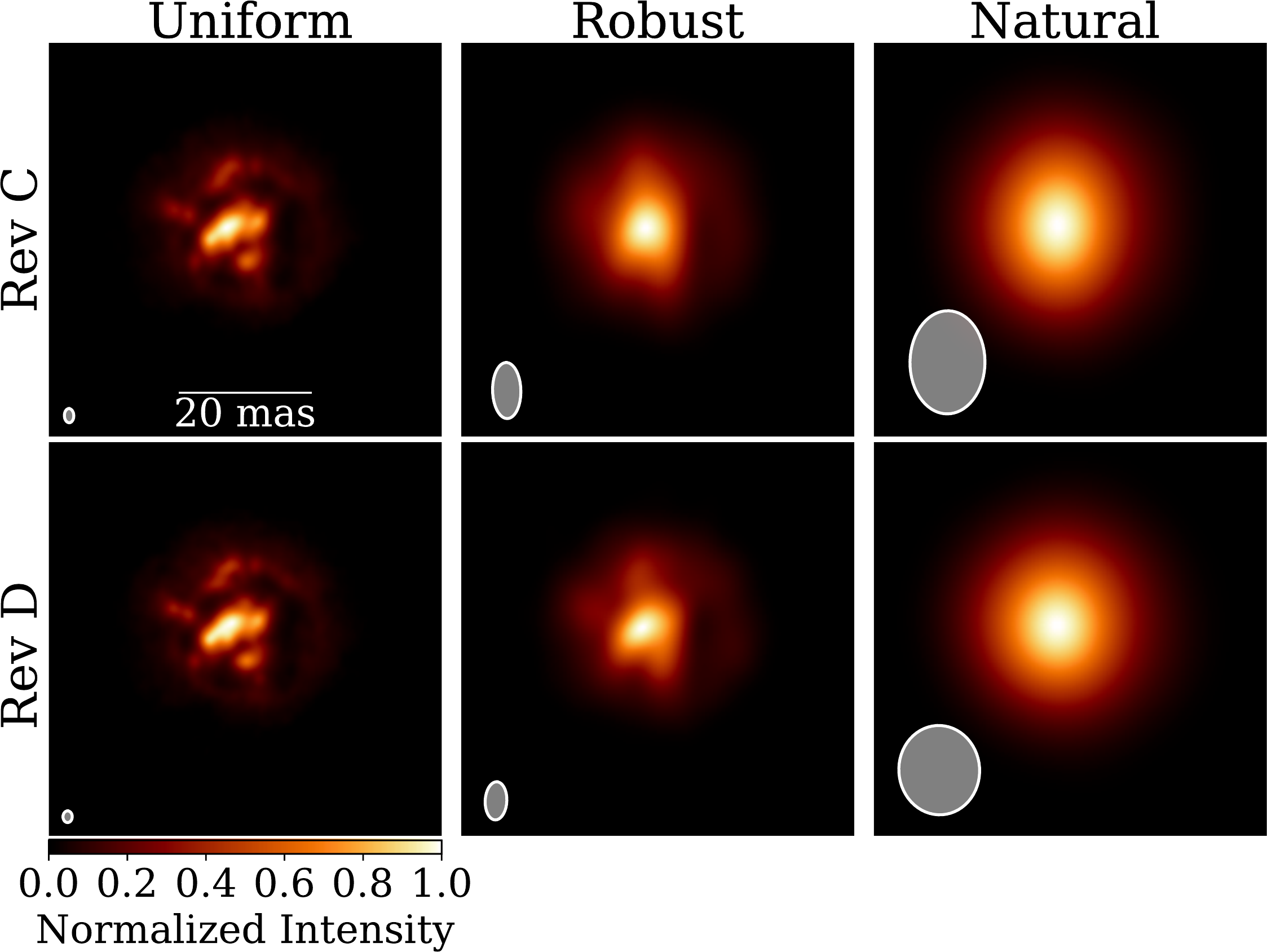}{0.5\textwidth}{(a) {\tt Freytag} Model}
        \fig{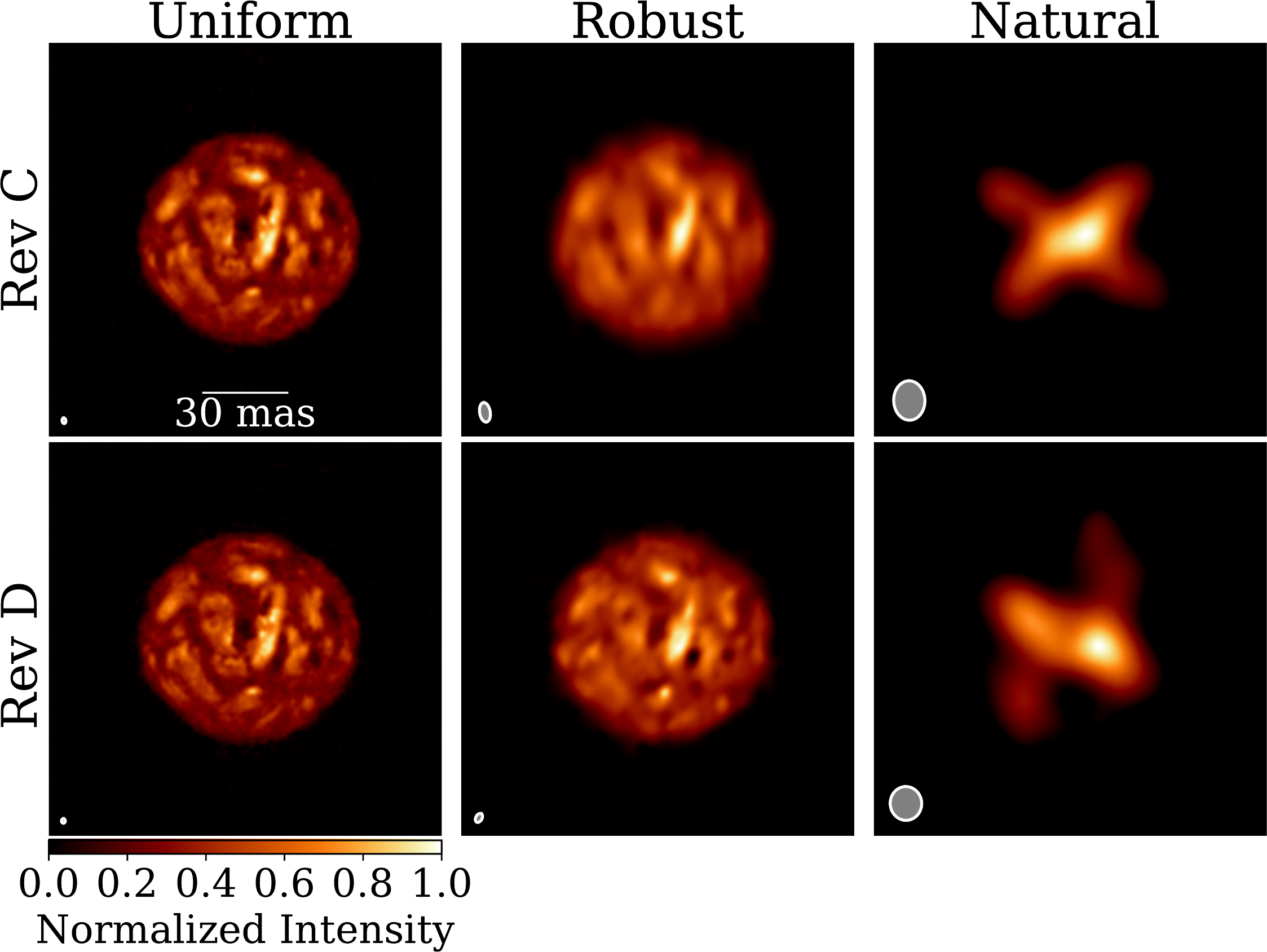}{0.5\textwidth}{(b) {\tt Chiavassa} Model}
    }
    \gridline{
        \fig{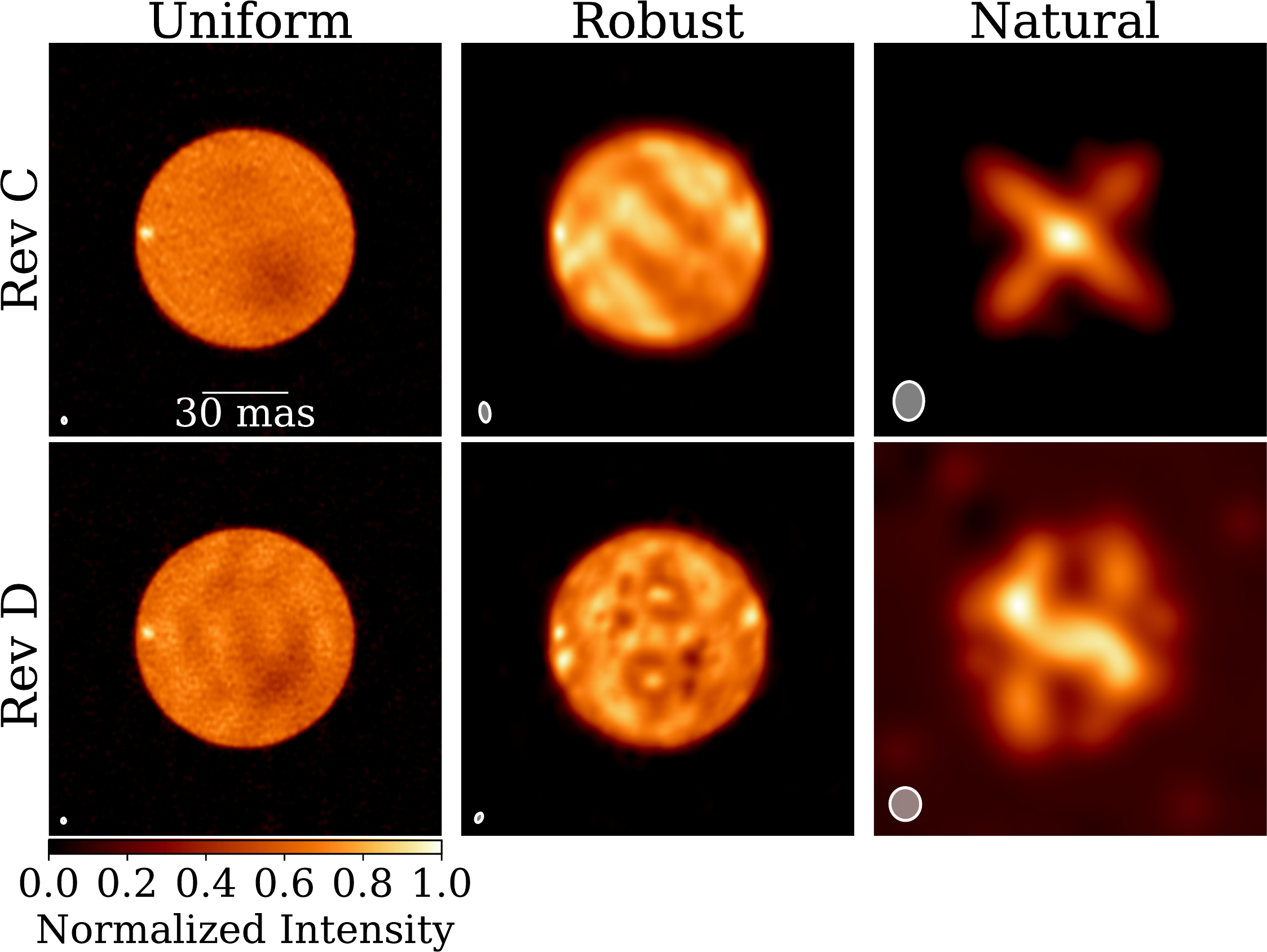}{0.5\textwidth}{(c) {\tt RSD222PC} Model}
        \fig{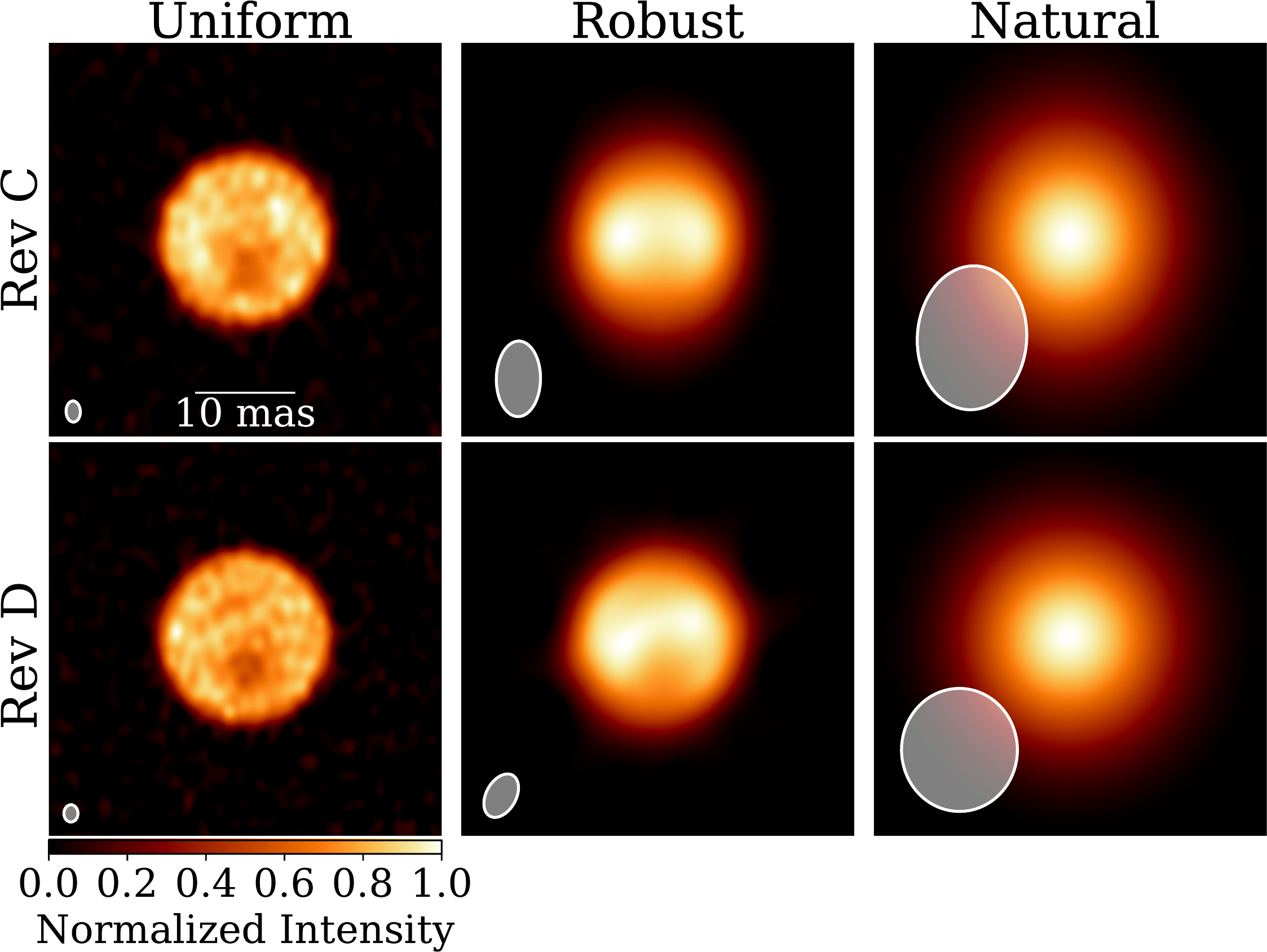}{0.5\textwidth}{(d) {\tt RSD1KPC} Model}
    }
    \caption{\msclean reconstructions at different $uv$-weightings for both array configurations.}
    \label{fig:clean_rec}
\end{figure*}

\begin{figure*}[ht]
    \centering
    \gridline{
        \fig{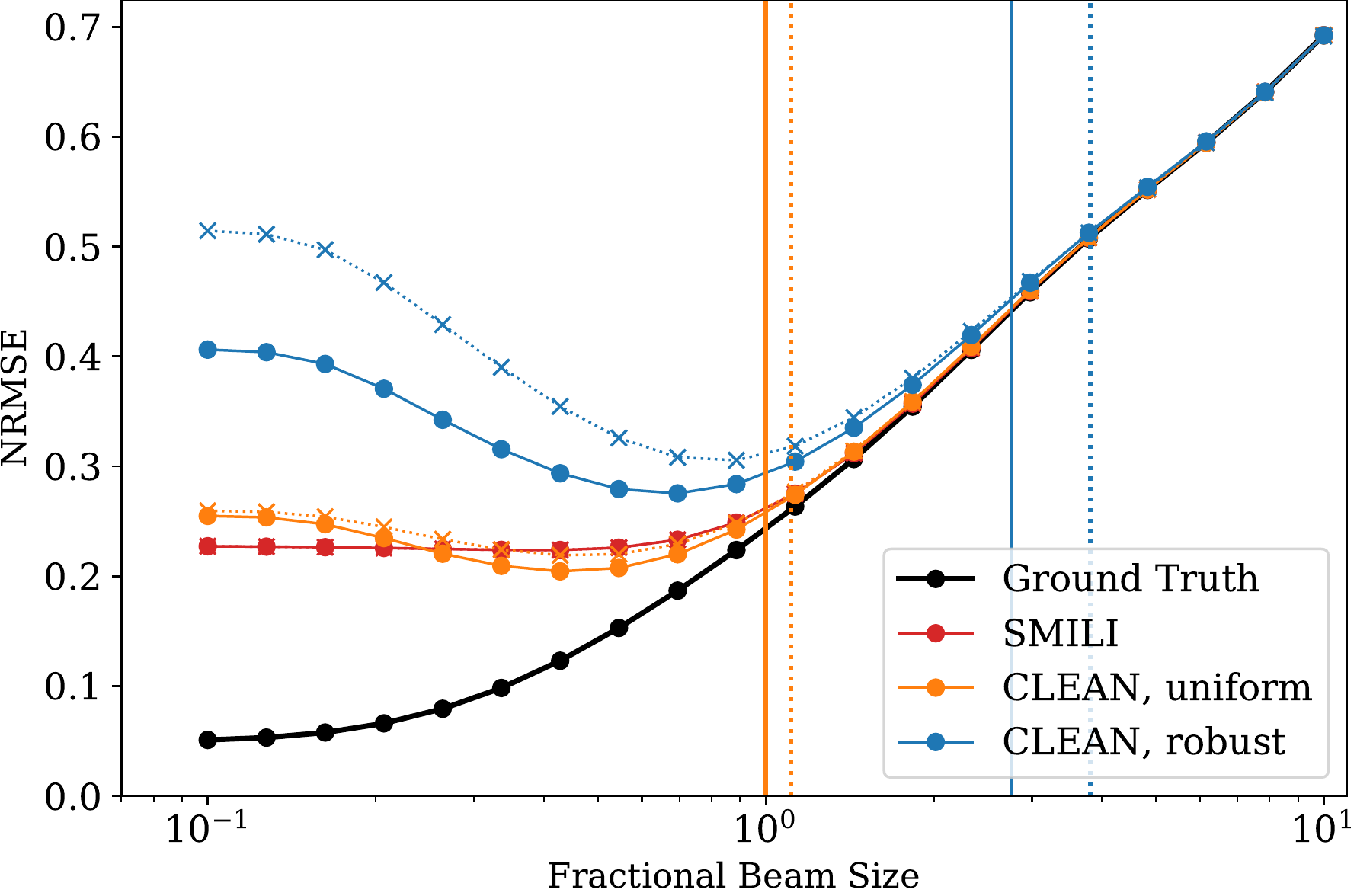}{0.5\textwidth}{(a) {\tt Freytag} Model}
        \fig{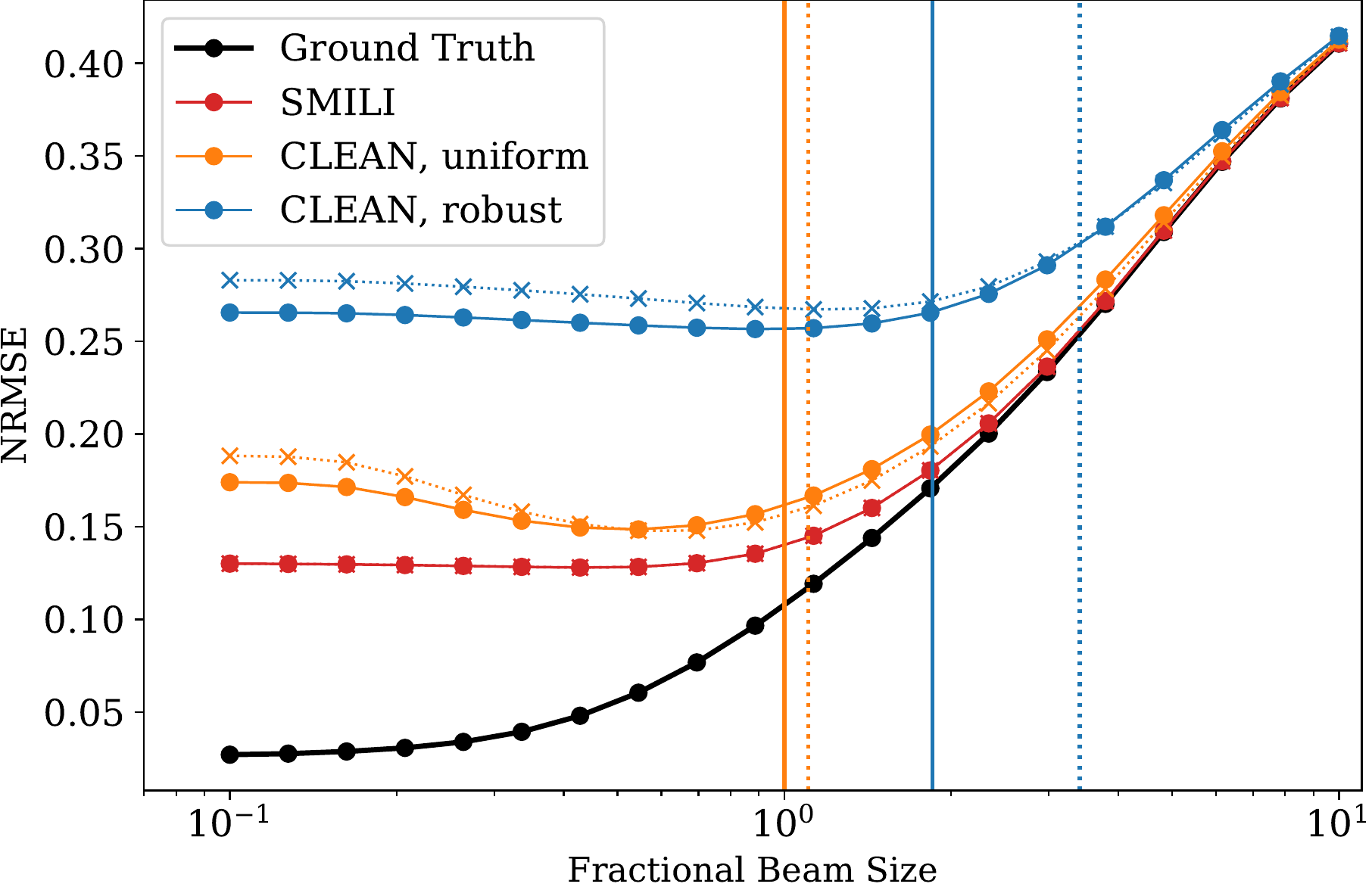}{0.5\textwidth}{(b) {\tt Chiavassa} Model}
    }
    \gridline{
        \fig{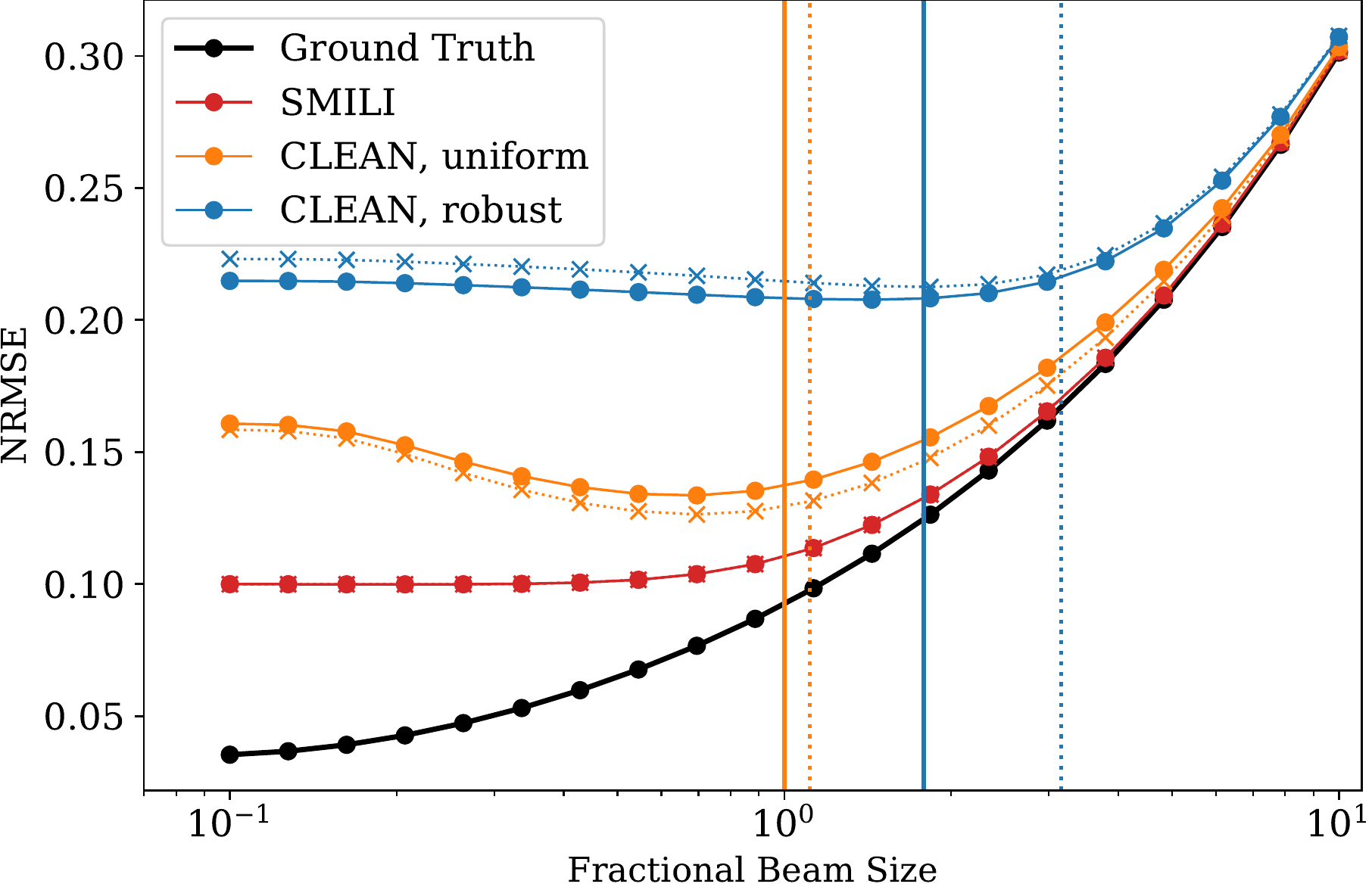}{0.5\textwidth}{(c) {\tt RSD222PC} Model}
        \fig{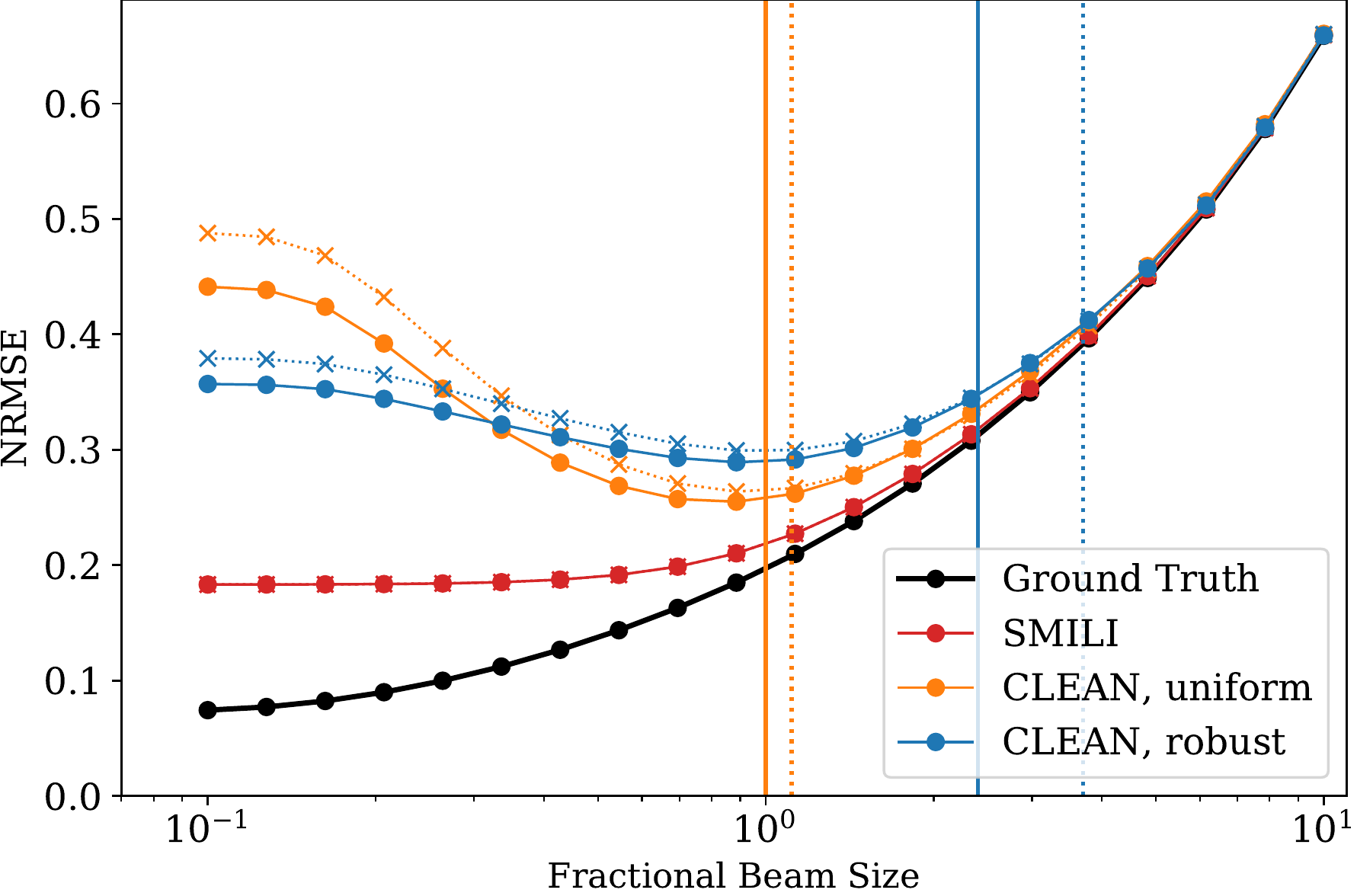}{0.5\textwidth}{(d) {\tt RSD1KPC} Model}
    }
    \caption{The normalized root-mean-square errors (NRMSEs) of reconstructions as a function of the restoring beam size. Each NRMSE curve was calculated between the corresponding beam-convolved image and the {\it non-convolved} groundtruth image adopted as the reference. The beam size on the horizontal axis is normalized to that of uniform weighting of the \revd configuration used in \casa imaging. The solid lines show the NRMSE curves for the \revd configuration, while the dotted lines show those for the \revc configuration. Each vertical line indicates the angular scale giving the same beam solid angle for the array configuration with the corresponding line style as the $uv$-weighting of the corresponding line color.}
    \label{fig:nrmse}
\end{figure*}

 Figure~\ref{fig:rml_imaging} shows the groundtruth images and the \texttt{SMILI} reconstructions, which are convolved with the same Gaussian beam for the \revd configuration in uniform weighting\footnote{RML methods often provide piece-wise smooth images, even without blurring the beam, which may provide high fidelity images at a modest superresolution. Please see our previous memo \citep{Akiyama2019} for the unblurred groundtruth images and RML reconstructions.} to match the resolution across the groundtruth and reconstructions from different array configurations. Comparing the results of imaging using both Rev\,C and Rev\,D array configurations, there are not many noticeable differences for the \texttt{Freytag}, \texttt{Chiavassa} and \texttt{UniDisk222pc} models.
The \texttt{UniDisk222pc} model appears slightly more uniform in shape for the rev\,D reconstructions when looking at the North-Western and South-Eastern edges of the star.
For the \texttt{UniDisk1kpc} model, the Rev\,D reconstruction shows slightly better reconstruction with a more uniform brightness and better localization of a bright spot at the Eastern edge of the disk.

Figure~\ref{fig:clean_rec} shows the \texttt{CASA} reconstructions for each stellar model with both array configurations using uniform, robust, and natural weightings. Overall, the natural weighting reconstructions lose most, if not all, of the compact stellar structure for both arrays. 
Although the loss in the angular resolution is expected by its definition, for both {\tt Chiavassa} and {\tt RSD222PC} models, where the structure is more extended than the fitted beam size, natural weighting does not allow the reconstruction of a uniform circular disk and instead artificially creates an X-shape structure.
For robust weighting, the \revd reconstructions provide clearly better reconstructions across all four models at finer resolutions than the \revc configuration. The uniform weighting reconstructions do not provide substantial differences for all four models. 

For a more quantitative analysis with matching the resolution of images from different techniques and different array configurations, in Figure \ref{fig:nrmse} we show characteristic levels of reconstruction errors at each spatial scale using the normalized root-mean-square error \citep[NRMSE;][]{Chael2016} for both array configurations and all reconstructions except for \msclean images at natural weighting. NRMSE is defined by
\begin{equation}
    {\rm NRMSE}({\bf I},\,{\bf K}) = \sqrt{\frac{\sum_i |I_i - K_i|^2}{\sum_i |K_i|^2}},
\end{equation}
where ${\bf I}$ is the image to be evaluated, and ${\bf K}$ is the reference image. We adopt the non-convolved groundtruth image as the reference image, and evaluate NRMSEs of the groundtruth and reconstructed images convolved with an elliptical Gaussian beam equivalent to the one appropriate for uniform weighting with the \revd configuration. The curve for the groundtruth image shows the errors caused by the limited angular resolution. As shown by the vertical lines, the improvement in the beam size with the \revd configuration is more significant with the robust weighting. 

Figure \ref{fig:nrmse} shows that, even after matching resolutions, the \revd configuration provides reconstructions with lower NRMSE for robust weighting, suggesting that CLEAN components are well modeled with the \revd configuration. However, the relative improvement at the same resolution is less than that provided by the improvement in the angular resolutions, which one can see by taking differences in NRMSE values at the cross points of the vertical lines and NRMSE curves. This indicates that the \revd's improvement in the image appearance shown in Figure \ref{fig:clean_rec} for robust weighting is primarily attributed to its finer beam size. For uniform weighting, the \revc and \revd reconstructions do not give substantial differences, even at the same resolution or even considering the slight improvement in the beam size.

RML reconstructions with {\tt SMILI} show almost identical NRMSE curves for all models, which is consistent with the visual appearances shown in Figure \ref{fig:rml_imaging}. These demonstrate resiliency of the performance for different array configurations. Consistent with our previous work \citep{Akiyama2019}, RML reconstructions with {\tt SMILI} outperform \texttt{MS-CLEAN} reconstructions with {\tt CASA} for a wide range of spatial scales, including the nominal resolution at uniform weighting. An exception is the {\tt Freytag} model with its many compact emission features. 

\section{Summary}
We have presented a study exploring stellar imaging with the ngVLA, continued from the study of \cite{Akiyama2019}. In this memo, we evaluated the imaging performance with an updated version of the Main Array configuration (\revd) and of \msclean imaging at different $uv$-weightings. Here are the main results of this work:
\begin{enumerate}
    \item The \revd configuration provides better $uv$-coverage, resulting in a modest improvement in the synthesized beam regardless of the choice of $uv$-weighting. The synthesized beam has a better circular symmetry and a lower tail of side lobes. 
    \item Both RML and uniform-weighted \msclean reconstructions provide high-fidelity reconstructions for all stellar models adopted in this work with both the \revd and the previous \revc configurations, demonstrating that the capability of stellar imaging does not significantly depend on the configurations of the Main Array.
    \item For \msclean imaging, the \revd configuration provides visually noticeable improvement in the recovery of the complex morphology in detailed stellar models when adopting robust weighting. The NRMSE analysis with the matched resolutions indicates that this is primarily attributed to the finer restoring beam rather than the better modeling of CLEAN components. 
    \item Stellar imaging with \msclean using natural weighting is severely limited by the highly non-Gaussian beam, and this problem remains with the \revd configuration. The highly non-Gaussian nature of the synthesized beam still persists with both robust and natural weightings in the \revd configuration, and may induce artifacts in the \msclean reconstructions, even if the source structure is larger than the corresponding angular resolution.
    It may severely limit the capability of natural weighting for what it was traditionally considered to enhance --- for instance, the fidelity of imaging more extended emission from a surrounding nebula or circumstellar material around a star, or the accurate recovery of the total flux for a given angular resolution. 
    \item RML images are almost identical between the \revd and previous \revc configurations for all four stellar models considered here, demonstrating the resiliency of its performance over different configurations. The RML methods are shown to have better or comparable performances than \msclean in the presented simulations, consistent with our previous work presented in \citet{Akiyama2019}.
\end{enumerate}
Our simulations imply that the \revd configuration will provide a better imaging capability compared with previously proposed Main Array configurations, and they continue to demonstrate that RML methods are an attractive choice for imaging, even for the improved array configuration. 

We note that the improvement of imaging capability provided by the \revd configuration may be more significant for more realistic cases with residual calibration errors and where the better synthesized beam may be helpful in distinguishing true signals from noise while {\tt CLEAN}ing. In future work, we will further study more realistic cases with residual calibration errors and examine spectral line imaging capabilities for the \revd configuration as outlined in \citet{Akiyama2019}.

\section*{Acknowledgments} \noindent
We thank Chris Carilli, Brian Mason and Eric Murphy for sharing the \revd configuration, which motivates this work. 
This work was supported by the ngVLA Community Studies program, coordinated by the National Radio Astronomy Observatory (NRAO), which is a facility of the National Science Foundation (NSF) operated under cooperative agreement by Associated Universities, Inc.
Developments of \texttt{SMILI} at MIT Haystack Observatory have been financially supported by grants from the NSF (AST-1440254; AST-1614868; AST-2034306).
The Black Hole Initiative at Harvard University is financially supported by a grant from the John Templeton Foundation. Special thanks to the Barry Goldwater Scholarship and Excellence in Education Foundation.

\end{document}